%% file: main.tex
\documentclass[journal,twoside]{IEEEtran} 
\usepackage{amsfonts}
\usepackage{float}
\usepackage{comment}
\usepackage{booktabs}
\usepackage{fancyvrb}
\usepackage{graphicx}
\usepackage[dvipsnames]{xcolor}
\usepackage{framed}
\usepackage{todonotes}
\definecolor{shadecolor}{rgb}{.95, .95, .95}
\newenvironment{code}%
   {\snugshade\Verbatim}%
   {\endVerbatim\endsnugshade}
\usepackage{amsmath}
\usepackage{algorithm}
\usepackage{mathtools}
\usepackage{algorithmicx} %
\usepackage[noend]{algpseudocode} %

\usepackage{color} 
\usepackage{soul}

\usepackage{caption}
\usepackage{subcaption}

\usepackage{ulem} 
\definecolor{darkgreen}{cmyk}{0.90,0,0.80,0.20}
\definecolor{darkred}{cmyk}{0, 1, 1, 0.10}
\definecolor{lightblue}{cmyk}{1, 0.70, 0, 0}

\usepackage{array}
\newcolumntype{x}[1]{>{\centering\arraybackslash\hspace{0pt}}p{#1}}

\DeclareMathOperator*{\argmax}{arg\,max}

\begin{document}

\title{Fault Injection on Embedded Neural Networks: Impact of a Single Instruction Skip}

\author{
\IEEEauthorblockN{Clément Gaine\IEEEauthorrefmark{1}, Pierre-Alain Moëllic\IEEEauthorrefmark{2}\IEEEauthorrefmark{3}, Olivier Potin\IEEEauthorrefmark{1},  Jean-Max Dutertre\IEEEauthorrefmark{1}}

\IEEEauthorblockA{\IEEEauthorrefmark{2}CEA Tech, Centre CMP, Equipe Commune CEA Tech - Mines Saint-Etienne, F-13541 Gardanne, France}
\IEEEauthorblockA{\IEEEauthorrefmark{3}Univ. Grenoble Alpes, CEA, Leti, F-38000 Grenoble, France \\ 
\emph{pierre-alain.moellic@cea.fr}}

\IEEEauthorblockA{\IEEEauthorrefmark{1}Mines Saint-Etienne, CEA, Leti, Centre CMP, F-13541 Gardanne, France \\
\emph{c.gaine@emse.fr, olivier.potin@emse.fr, dutertre@emse.fr}}
}


\markboth{Preprint version, accepted at DSD2023, AHSA special session}{}

\maketitle

\begin{abstract}
\input{src/abstract}

\end{abstract}

\section{Introduction}
\input{src/introduction}

\section{Background \& Related works}
\input{src/sota_position}

\section{Experimental setup}
\input{src/experimental_setup}

\section{Instruction skip on a CNN inference}
\label{exp_section}
\input{src/results}

\section{Discussions}
\label{discussions}
\input{src/discussions}

\section{Conclusion}
\input{src/conclusion}

\section*{Acknowledgment}
This work is supported by (CEA-Leti) the European project InSecTT (ECSEL JU 876038) and by the French ANR in the \textit{Investissements d’avenir} program (ANR-10-AIRT-05, irtnanoelec);  and (MSE) by the ANR PICTURE program. This work benefited from the French Jean Zay supercomputer with the AI dynamic access program.

\bibliographystyle{splncs04}
\bibliography{biblio}

\end{document}

%% file: src/abstract.tex
With the large-scale integration and use of neural network models, especially in critical embedded systems, their security assessment to guarantee their reliability is becoming an urgent need. More particularly, models deployed in embedded platforms, such as 32-bit microcontrollers, are physically accessible by adversaries and therefore vulnerable to hardware disturbances. We present the first set of experiments on the use of two fault injection means, electromagnetic and laser injections, applied on neural networks models embedded on a Cortex M4 32-bit microcontroller platform. Contrary to most of state-of-the-art works dedicated to the alteration of the internal parameters or input values, our goal is to simulate and experimentally demonstrate the impact of a specific fault model that is \textit{instruction skip}.
For that purpose, 
we assessed several modification attacks on the control flow of a neural network inference.  
We reveal integrity threats by targeting several steps in the inference program of typical convolutional neural network models, which may be exploited by an attacker to alter the predictions of the target models with different adversarial goals.    

%% file: src/introduction.tex
Security of Machine Learning (ML) models is one of the most important challenge of modern Artificial Intelligence, amplified by the massive deployment of models (more particularly neural networks) in a large variety of hardware platforms. Those platforms include devices with strong constraints in terms of memory, computing ability, latency or energy (e.g., for IoT-oriented applications). The \textit{adversarial} and \textit{privacy-preserving} ML communities have already demonstrated an impressive set of threats that target the integrity, confidentiality and availability of models~\cite{papernot2018sok}. However, most of these attacks can be referred as \textit{theoretical} or \textit{algorithmic} since they consider a model as an \textit{abstraction} and do not rely on the specific features of their software or hardware implementations. Most recently, the attack surface has been significantly widened with such \textit{implementation}-based attacks that leverage software or hardware characteristics as well as theoretical backgrounds highlighted by previous attacks. It is the case for \textit{weight-based adversarial attacks} such as the Bit-Flip Attack (BFA)~\cite{rakin2019bit} that directly disturbs the internal parameters of a deep neural network model stored in memory (typically, DRAM or Flash). Interestingly, in the BFA, the selection of the most sensitive parameters follows a gradient-based approach similar to classical white-box \textit{adversarial examples} crafting methods. This leads to only a few bit-flips to drop the accuracy of a state-of-the-art convolutional neural network to a random-guess level. Another example, is the use of side-channel analysis \cite{joud2023practical} or fault injection attacks (as rowhammer in \cite{rakin2022deepsteal}), to totally or partially recover the values of parameters so that it could significantly increase the efficiency of a \textit{model extraction} attack that aims at stealing a black-box protected model. 

Except for passive side-channel analysis, most of these new implementation-based threats are data-oriented fault injection attacks targeting the stored parameters. In this work, we highlight another important attack vectors caused by fault injection that target the instruction flow, more particularly with \textit{instruction skips}. To the best of our knowledge, this work is the first to demonstrate instruction skip with laser and electromagnetic fault injection in the inference of neural network models deployed in a Cortex-M 32-bit microcontroller. 

\textbf{Our contributions} are the followings:
\begin{itemize}
    \item We used two injection means on a Cortex-M4 platform, electromagnetic and laser injections, and target the inference of a standard convolutional neural network performing an image classification task.  
    \item We demonstrate and analyze the impact of a single instruction skip at different critical paths of the inference: convolutional layers, bias additions, activation functions.
    \item We highlight two potential adversarial exploitation: memory effect and forced prediction.
\end{itemize}

%% file: src/sota_position.tex
\subsection{Background}
\noindent\textbf{Neural network models.} A supervised neural network model $M_\Theta(x)$ is a parametric model trained to optimally map an input space $\mathcal{X}=\mathbb{R}^n$ (e.g., images) to an output space $\mathcal{Y}$.  
For a classification task, $\mathcal{Y}$ is a finite set of labels $\{1,...,C\}$. The neural network model $M_\Theta : \mathcal{X} \rightarrow \mathcal{Y}$, with parameters $\Theta$ (also referred as \textit{weights}), classifies an input $x \in \mathcal{X}$ to a set of raw or normalized scores in $\mathbb{R}^{C}$ so that the predicted label is $ \hat{y} = \argmax(M_\Theta(x))$.
$M_\Theta$ is trained by minimizing a loss function $\mathcal{L}\big(M_\Theta(x),y\big)$ that quantifies the error between the prediction $M_\Theta(x)$ and the \textit{groundtruth} $y$. The training process aims at finding the best parameters that minimize the loss on the training dataset.

\noindent\textbf{A Perceptron} (also called~\textit{neuron}) is the basic functional element of a neural network. It first processes a weighted sum of the input $x=(x_0,..., x_j, x_{n-1}) \in \mathbb{R}^n$ with its trainable parameters $\theta$ and $b$ (called  \textit{bias}), then it non-linearly maps the output thanks to an \textit{activation function} $\sigma$: $a(x)=\sigma( \theta_{0}x_0+...+\theta_{n-1}x_{n-1}+b)$, where $a$ is the perceptron output. A classical activation function is the rectified linear unit (ReLU) defined as $\sigma(x)=max(0,x)$. 

\noindent\textbf{MultiLayer Perceptron (MLP)} are  deep neural networks composed of several vertically stacked neurons called \textit{layers}. These layers are called \textit{fully-connected} or \textit{dense} or even \textit{linear}. For a MLP, a neuron from layer $l$ gets information from all neurons belonging to the previous layer $l-1$, therefore the output of a neuron is defined as in~\ref{eq_mlp}:
\begin{equation}
    a^{l}_j(x)=\sigma\Big(\sum_{i \in (l-1)}{\theta_{i,j}}a^{l-1}_i + b_j\Big)
\label{eq_mlp}
\end{equation}
\noindent where $\theta_{i,j}$ is the weight that connects the $j^\text{th}$ neuron of the $l^\text{th}$ layer and the $i^\text{th}$ neuron of the previous layer ($l-1$), $b_j$ is the bias of neuron $j$ of layer $l$ and $a^{l-1}_i$ and $a^{l}_j$ are the outputs of neuron $i$ of layer $(l-1)$ and neuron $j$ of layer $l$, respectively.

\noindent\textbf{Convolutional Neural Network (CNN)} is another type of neural network models that used convolutions with a set of \textit{kernels} (also called \textit{filters}). The trainable weights are the parameters of the kernels and are shared among the input. The kernels are usually square with low dimensions, typically 3x3 for image classification. Therefore, for a \textit{convolutional layer} composed of $K$ kernels of size $Z$ applied on an input tensor of size $H\times W \times C$, the weights tensor $\Theta$ will have the shape $[K,Z,Z,C]$ (i.e., $KCZ^2$ parameters without bias, $(K+1)CZ^2$ otherwise). A naive implementation of the convolution of an input tensor $X$ and a set of $K$ kernels is detailed in algorithm~\ref{algo_naive_conv}.
\begin{algorithm}
 \normalsize
\caption{Convolution layer ($K$ kernels)}\label{algo_naive_conv}
\textbf{Input:} Tensor $X$ of size $H\times H\times C$, parameters tensor $\Theta$ of size $Z\times Z\times C\times K$, bias tensor of size $K$\\
\textbf{Output:} Tensor $Y$ of size $H\times H\times K$
\begin{algorithmic}[1]
\For{$k$ in $[1,K]$}
    \For{$x$ in $[1,H]$}
        \For{$y$ in $[1,H]$} 
            \State $Y_{x,y,c}= B_{k}$
            \For{$m$ in $[1,Z]$}
                \For{$n$ in $[1,Z]$}
                    \For{$c$ in $[1,C]$} 
                        \State $Y_{i,j,c}+= \theta_{m,n,k,c} \cdot X_{x+m,y+n,k}$
                    \EndFor
                \EndFor
            \EndFor
        \EndFor
    \EndFor
\EndFor
\Return $Y$
\end{algorithmic}
\normalsize
\end{algorithm}

The output is also mapped with an activation function such as ReLU. Then, a third operation is applied with a \textit{pooling} process that aims at reducing the dimensions of the output tensor by locally summing it up with some statistics. A classical approach is to apply a \textit{Max pooling} or an \textit{Average pooling} with a 2x2 kernel over the output tensor $Y$ of size $H\times H\times K$ so that the resulting tensor is half the size $(H/2)\times (H/2)\times K$. Pooling also provides interesting (small) translation invariance property.   

\noindent\textbf{Embedded models.} For a typical 32-bit microcontroller, the model parameters are stored in the Flash memory and internal computations (i.e., mainly multiply-accumulations and non-linear activation) are processed in SRAM. To embed complex ML models and fit the memory and latency requirements, classical compression techniques encompass model pruning~\cite{zhu2018prune} and  quantization~\cite{zhou2016dorefa, courbariaux2015binaryconnect}. 
For 32-bit MCU, 8-bit quantization of the weights is a standard performed as a post-processing step (after training) or at training step with training-aware quantization methods. Post-training 8-bit quantization is proposed as the default configuration in many deployment tools (e.g., TF-Lite, CubeMX.AI, NNoM, MCUNet) and may be applied for both weigths and activation outputs.

We used the NNoM (Neural Network on Microcontroller)\footnote{https://majianjia.github.io/nnom/} deployment framework, an open-source library with a full access to the source code (C) that enables 8-bit quantization for the weights, biases, activation values and output scores. The quantization is performed in the same way as in ARM CMSIS-NN~\cite{lai2018cmsis} and relies on a uniform symmetric powers-of-two scheme (Eq.~\ref{eq_powers_of_two}) that avoid division operation with only integer additions, multiplications and bit shifting.

\begin{equation}
   x_{i} = \left \lfloor{x_{f}\cdot 2^{7-dec}}\right \rceil \text{,   } dec = \left \lceil{log_{2}\big(max\big(|X_{f}|\big)\big)}\right \rceil
\label{eq_powers_of_two}
\end{equation}

where $X_{f}$ is a 32-bit floating point tensor, $x_{f}$ a value of $X_f$, $x_{i}$ its 8-bit counterpart and $2^{7-dec}$ the quantization scale.

\noindent\textbf{Fault injection attacks} (FIA) are active hardware threats~\cite{Barenghi2012} that usually require a physical access to the victim device~\cite{breier_sigsac}. 
Fault injection techniques gather global approaches such as voltage or clock glitching  and moderate/high-cost methods such as laser (LFI) or electromagnetic (EMFI) that reach high temporal and spatial accuracy~\cite{agoyan2010flip}. EMFI involves generating a magnetic field that causes voltage variations in the circuit, leading to alter propagation times of the signals through logic gates. This can result in faults where assembly instruction are modified or skipped.
For LFI, a laser diode emits photons that create a photocurrent when they reach the sensitive points of the targeted microcontroller, resulting in voltage variations. This can change bit values, leading to instruction opcode modifications, which can transform one instruction into another or the \texttt{nop} instruction. In this case, we obtain an \textit{instruction skip} similar to those obtained with EMFI. Interestingly,  even though these two methods use different physical mechanisms, the results can be similar.
Importantly, because of their precision and effectiveness, EMFI and LFI are standard fault injection means used in hardware security testing laboratories for security assessment or certification purposes~\cite{yuce2018fault}.   
\begin{figure*}[t!]
\centering
\includegraphics[width=0.64\textwidth]{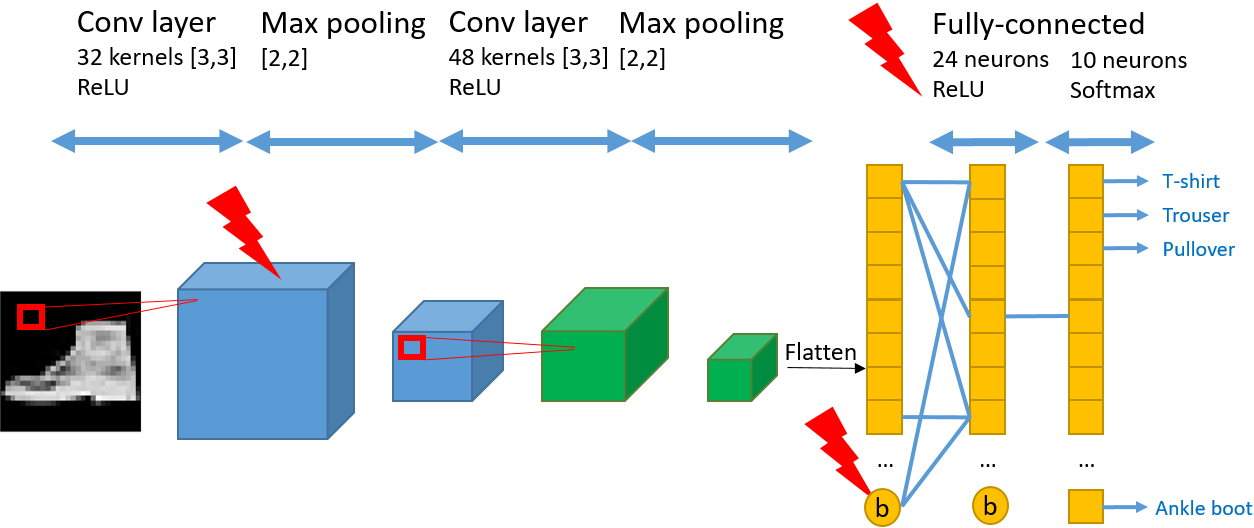}
\caption{Illustration of the CNN model. Red lightnings indicate the three investigated attack paths with instruction skip.}
\label{fig:attack_paths}
\end{figure*}
\subsection{Related works and positioning} 
Fault injection against deployed neural network models mainly focus on the alteration of the internal parameters stored in memory or the instructions flow. A reference for parameter-based attack is the Bit-Flip Attack (BFA)~\cite{rakin2019bit} that has been practically demonstrated with rowhammer on a DRAM platform~\cite{deephammer}.  
Typically, BFA successes in dropping the average accuracy of a state-of-the-art CNN model to a random-guess level with only a few tens of bit-flips. Rowhammer and bit-flips on the parameters have also been exploited for a model extraction scenario~\cite{rakin2022deepsteal} where the adversary knows the model architecture but has only access to less than 10\% of the training dataset. The attack first uses rowhammer (as in the rambleed attack~\cite{kwong2020rambleed}) to guess the value of the most significant bit of almost 90\% of the parameters of a victim model. Then, the adversary trains a \textit{substitute} model by constraining the weigths values with information previously extracted.

In the context of embedded neural network models, laser fault injection has been demonstrated on a 8-bit microcontroler (ATmega328P) by Breier et.al.~\cite{breier_sigsac} by targeting the instructions of some activation functions (ReLU, sigmoid, tanh implemented in C). Then, simulations on a 4-layer MLP trained on MNIST showed that it is necessary to perform a lot of faults (more than 50\% neurons faulted) on the last hidden layer to reach a reasonable attack success rate ($>50\%$). Other simulation works by Jap et.al.~\cite{jap} showed that a single bit modification on the Softmax activation function at the end of a neural network can lead to a misclassification. Liu et.al.~\cite{liu} achieved misclassification using clock glitch on a FPGA-based deep learning accelerator. Changing other physical parameters such as supply voltage can decrease accuracy. Salami et.al.~\cite{salami} demonstrated on FPGA-based CNN accelerators that in order to decrease the accuracy, it was necessary to reduce the supply voltage by at least 25\%.

To the best of our knowledge, our work is the first to demonstrate the impact of a single fault disrupting the instruction flow on the performance of a CNN model deployed in a Cortex-M platform. Contrary to~\cite{breier_sigsac} we consider a full inference program embedded with state-of-the-art deployment tool and analyze different attack paths. Additionally, with this scope, our experiments are the first to demonstrate both electromagnetic and laser injections for a complete inference program (CNN trained on the standard Fashion MNIST dataset) on a 32-bit ARM Cortex-M4 platform.

%% file: src/experimental_setup.tex
\subsection{Device under test, model and dataset}
We used a 32-bit ARM Cortex-M4 microcontroller as target which can operate at a frequency of up to 100\,MHz. The device has a 512KB Flash memory and a RAM of 128KB.

We focused our experiments on a typical convolutional neural network model trained for a supervised image classification task. We used the standard Fashion-MNIST dataset, which consists of 70,000 (60K for training and 10K for testing) 28x28 grayscale images divided into 10 cloth categories. Our model is composed of two convolutional layers with respectively 32 and 48 kernels of size $[3,3]$ with ReLU as activation. Each layer is followed by a Max pooling layer of size $[2,2]$. The end of the model is composed of two fully-connected layer with respectively 24 and 10 neurons. The activation function is ReLU except for the last layer which typically used Softmax to provide normalized outputs. 
The model (illustrated in Fig~\ref{fig:attack_paths}) has a total of 70,914 parameters and reaches an accuracy of 91\% on the complete test set.

The trained model (with TensorFlow v2) is deployed on our target device with the NNoM library~\cite{NNoM} that offers a 8-bit model quantization of the parameters, activation and output prediction scores and a complete white-box access to the inference code. The accuracy of the deployed 8-bit model is evaluated directly on the development board over limited random sets of 100 inputs. We observed that the implementation of the quantization scheme in NNoM raises integer overflow that may impact the accuracy depending on the test sets. Over different test sets the model has an accuracy from 77\% to 88\% (i.e., close to the accuracy of the full-precision model over the 10K test set). 
However, we noticed that our results are similar from one test sets to another (even by fixing the overflow issue). Therefore, for our fault injection experiments, we keep the same 100-input test set that reaches the lowest precision (77\%). 

\begin{figure}[h!]
\begin{center}
\includegraphics[width=0.96\columnwidth]{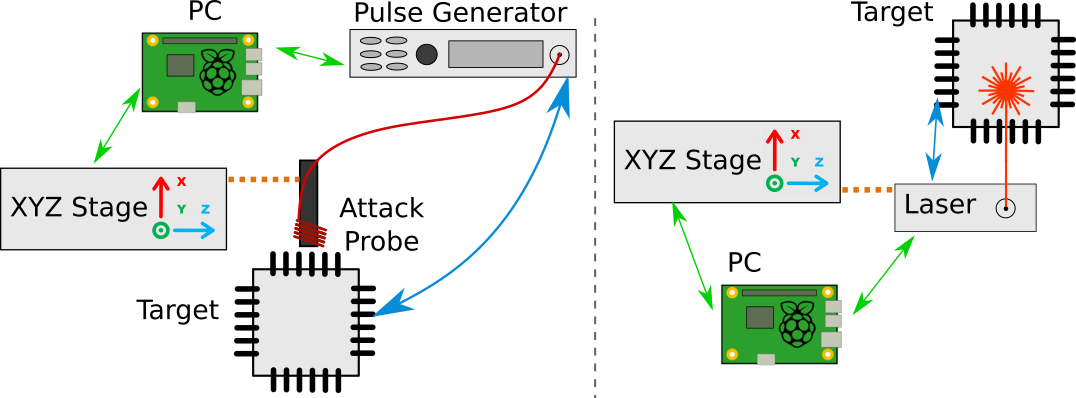}
\end{center}
\caption{EMFI (left) and LFI (right) setups.} \label{fig:em_et_laser_bench}
\end{figure}

\subsection{Fault injection setup}
For EMFI, we used a voltage pulse generator that can deliver 200\,V pulses, connected to an injection probe as illustrated in Figure~\ref{fig:em_et_laser_bench}. The control voltage of the pulse generator is 200\,V with a rise time of 2\,ns for a pulse width of 10\,ns. 

For LFI, we use an 1,064\,nm infrared laser beam with pulse energy of 0.1W, a pulse duration of 50\,ns and a spot size of 5\,µm. The laser attacks are performed on the rear face of the silicon, which requires to decapsulate the component. For the laser experiments, the operating frequency of the card is reduced from 100\,MHz to 50\,MHz.

The triggering of the laser and electromagnetic shot is synchronized by a signal generated by the target device and the delay between its rising edge and the triggering of the shot is adjustable. This enables to control the magnetic field and laser beam to target a specific instruction.

\subsection{Test code}

First, we need to map the sensitive areas of our device, that is to say the locations where exploitable faults are obtained. We used a simple assembly test code (Fig.~\ref{fig:code_registers}) to identify the locations where fault injections can be successfully performed. This code performs simple register manipulations and is long enough to not require precise time synchronization.
By comparing the readback values of the registers after execution of the test code with and without fault injection, we aimed at identifying a location where a \textit{sub} instruction is not executed. We successfully obtained an \textit{instruction skip} type fault.

\begin{figure}[h!]
\begin{code}[commandchars=\\\{\},fontsize=\scriptsize]
movs r3, #1
movs r4, #55
movs r5, #55
nop
...
nop
\textcolor{Red}{sub R4, R4, R3}
\textcolor{Red}{...}
\textcolor{Red}{sub R4, R4, R3}
\textcolor{Red}{sub R5, R5, R3}
\textcolor{Red}{...}
\textcolor{Red}{sub R5, R5, R3}
\textcolor{Green}{//readback of registers }
\end{code}
\caption{Test code manipulating registers}
\label{fig:code_registers}
\end{figure}

\begin{figure}[h!]
\begin{center}
\includegraphics[width=0.27\textwidth]{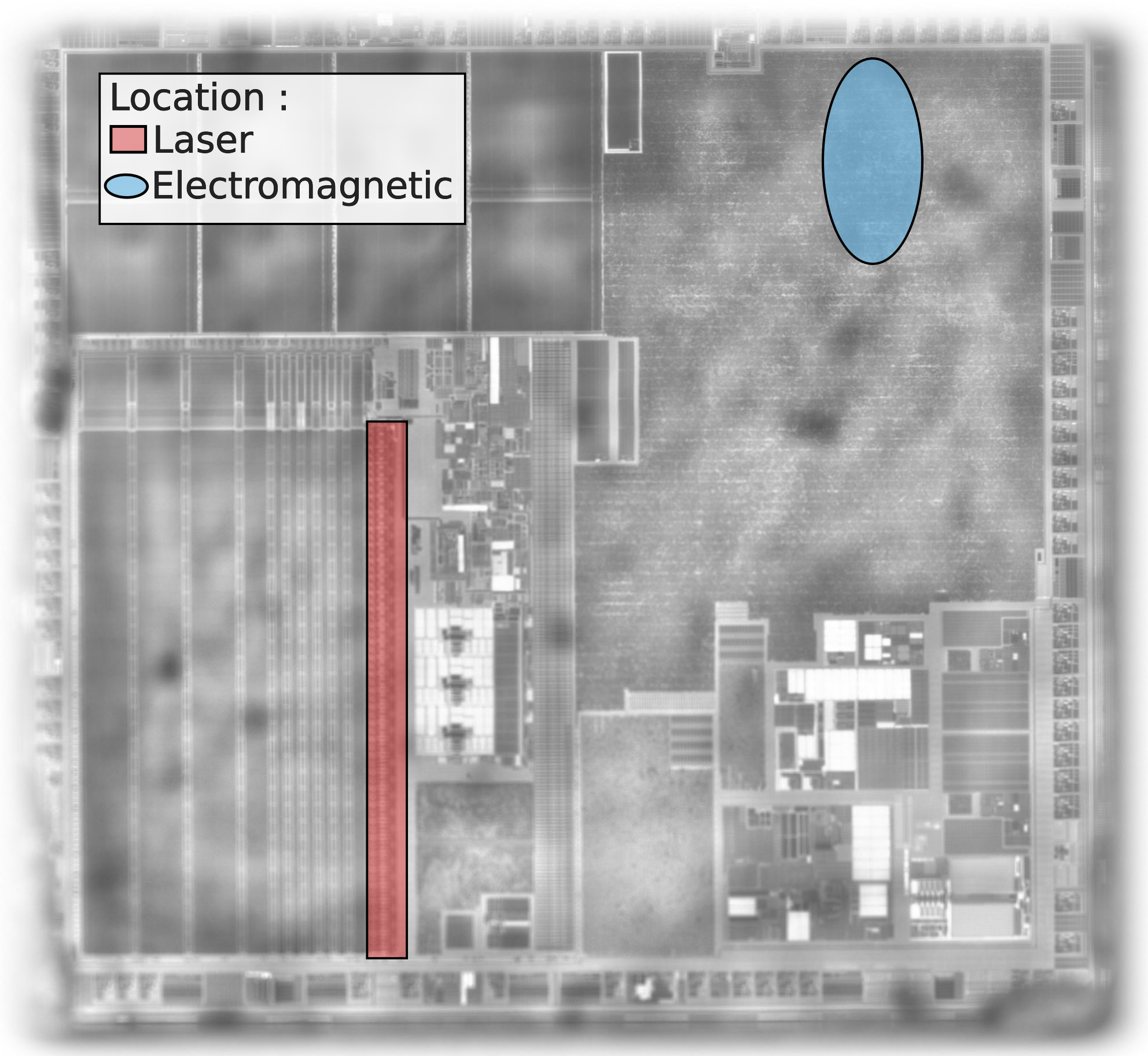}
\end{center}
\caption{Sensitive areas according to the different injection means.} 
\label{fig:imageir}
\end{figure}

With EMFI, the result of this mapping procedure is a 200\,µm by 100\,µm sensibility area within a chip size of 4\,mm by 4\,mm. It was necessary to use a very precise electromagnetic probe as described  in~\cite{wifs_emfi}. The positions of the probe for successful injections are indicated in blue on Figure~\ref{fig:imageir}. The positions used for LFI are distinct from those for EMFI and depicted in red in Figure~\ref{fig:imageir}. These results are consistent compared to other reference works such as \cite{DUTERTRE2021114133}.

%% file: src/results.tex
After characterizing the sensitive areas according to our fault model (instruction skip) and injection means (EM and laser pulse), we detail, in this section, three experiments on the inference program of our convolutional neural network model trained and tested on FashionMNIST. We targeted three attack paths that correspond to  critical elements of a CNN model: 
\begin{itemize}
    \item the first convolutional layer that extract low-level \textit{features} from the input;
    \item the bias addition and the ReLU activation function for the first fully-connected layer. 
\end{itemize}
An illustration of these attack paths on our CNN model is presented in Fig.~\ref{fig:attack_paths}.

\subsection{Targeting the first convolution layer}
\label{exp_conv}
\noindent\textbf{Experiments.} Parameter-based attacks such as the BFA~\cite{rakin2019bit} have highlighted the sensitivity of the first convolutional layers of CNN models against adversarial perturbations~\cite{hector2022closer}: an alteration of the initial features map grows and propagates through the network leading to a misprediction. However, this is achieved by targeting some specific kernels since others are resistant against the perturbation of their parameters. With our experiments, with only one instruction skip on the main convolution loop, we aimed at analyzing the sensitivity of this critical part of the inference.

During a convolution operation, a loop over the filters is executed to carry out the convolution computations (as in Algorithm~\ref{algo_naive_conv}). The assembly code of this loop is given in Fig.~\ref{fig:code_conv}. With an instruction skip, our objective is to prematurely interrupt the loop over the filters, thereby halting the execution of the convolutions. The impact of the instruction skip strongly depends of the implementation. In our case, such a fault completely breaks the convolution process: if a fault is injected for the kernel \textit{j} then the convolutions with kernels $[j+1;K]$ are not processed. We discuss that point in Section~\ref{discussions}.   

\begin{figure}
\begin{code}[commandchars=\\\{\},fontsize=\scriptsize]
\textcolor{Green}{//for (i=0, shift_idx=0; i<ch_im_out; }
\textcolor{Green}{      i++, shift_idx+=shift_steps)}
ldr   r3, [r7, #100]
adds  r3, #1
str   r3, [r7, #100]
ldr   r2, [r7, #72]
ldr   r3, [r7, #68]
add   r3, r2
str   r3, [r7, #72]
ldrh.w   r3, [r7, #132]
ldr   r2, [r7, #100]
cmp   r2, r3
\textcolor{Red}{blt.w   80037fa}
\end{code}
\caption{Assembly code of the loop over convolution filters.} \label{fig:code_conv}
\end{figure}

\begin{figure}[h!]
\centering
\includegraphics[width=0.45\textwidth]{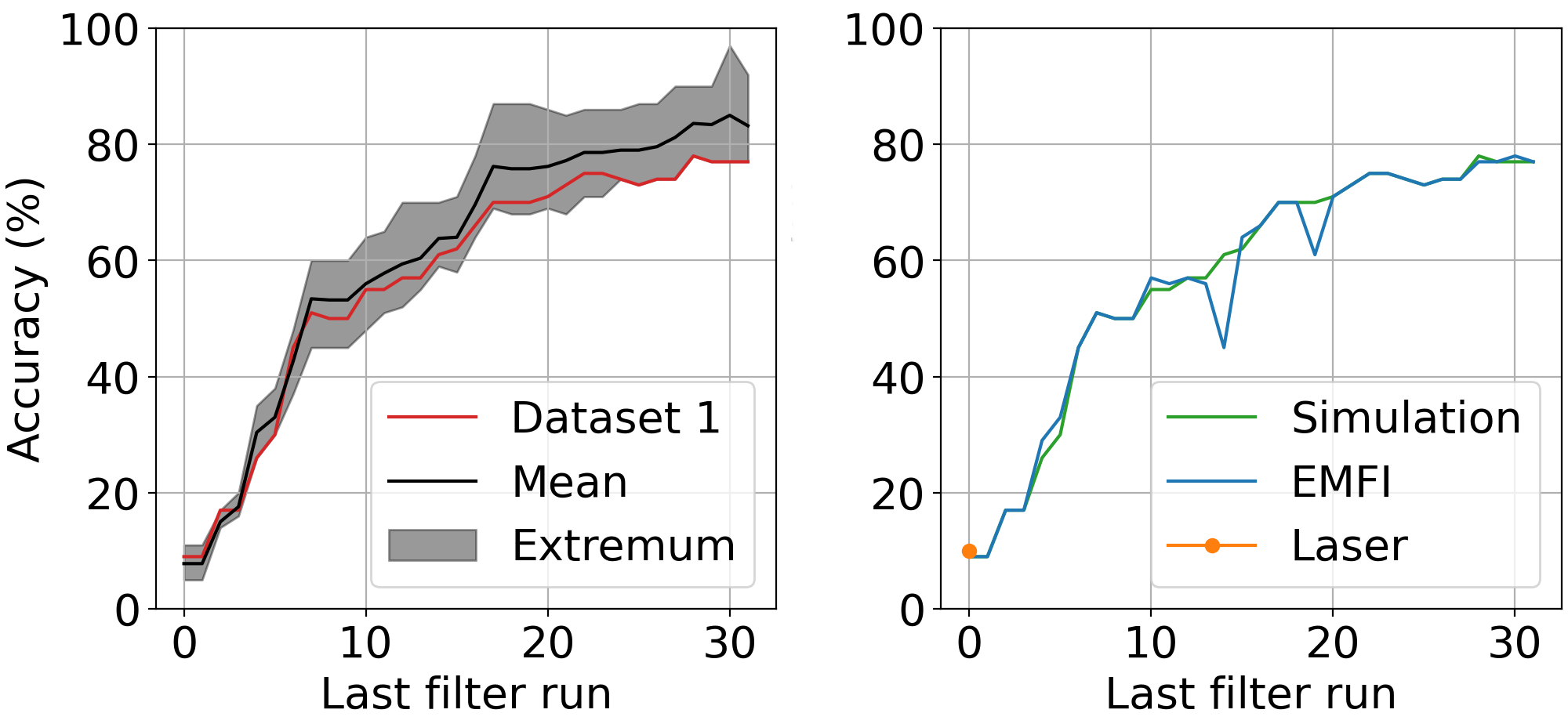}
\caption{ \textbf{a.}~Simulations over 5 different test sets (dataset used in \textbf{a} is refered as \textit{Dataset 1}) \textbf{b.}~Accuracy with an instruction skip on the first convolution layer (simulation and EMFI)} \label{fig:fashion_convo_jeux}
\end{figure}

Through simulation, we observed that a valid attack path is to jump the branch instruction (highlighted in red). Fig.~\ref{fig:fashion_convo_jeux}(a) reports the impact on the model accuracy of simulated fault injections (the skip of the jump instruction, hence ending the computation loop) as a function of the last processed filter index (the remaining filters were skipped).
We used five different random test sets of 100 images each. Despite minor variations, the results on each dataset are comparable. Thus, we can therefore consider that the 100 images of the first test set are representative of the general behavior (even if the accuracy is slightly below the average). 

Logically, the accuracy of the model decreases as the filters loop was exited earlier.
However, we observe that the last kernels do not have a significant impact: exiting the loop from the $17^{th}$ kernel~--~i.e. not performing almost half of the kernels~--~has a limited effect on the accuracy (from 75\% to 82\%). A possible explanation is that most of the deep neural network models are over-parametrized which can be observed when applying compression techniques such as \textit{model pruning} that typically remove, for standard CNN, a large part of kernels without  any significant drop of performance. Here, we can make the hypothesis that most of the useful features are \textit{captured} by the first kernels.            

To demonstrate this attack in practice, we conducted EMFI on the 100 images of the dataset 1. The experimental and simulation results, presented in Fig.~\ref{fig:fashion_convo_jeux} (b), are almost identical. However, since the repeatability of the EMFI is not perfect, it happens that the fault injection is not successful, which explains some results slightly above the simulation curve (accuracy is higher by a few percents). Other times, the EMFI will cause the board to crash, which occurs at filters 14 and 19 explaining the significant accuracy drops at these indexes.
For LFI, it appeared that triggering the position and the injection delay at different filters within the loop was very challenging. Therefore, we only processed one laser fault injection on the first filter to see if the LFI result was similar with our simulation and EMFI. We logically reached a random-guess level (10\%) since the forward pass is completely altered.

\begin{figure}[h!]
\centering
\includegraphics[width=0.97\columnwidth]{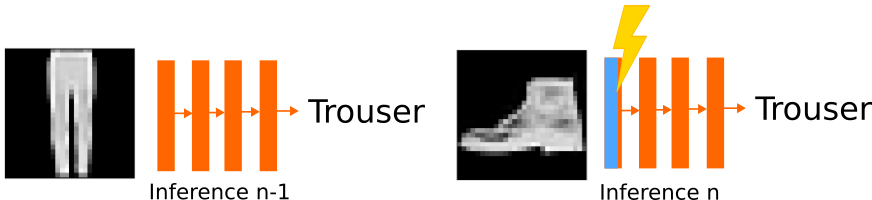}
\caption{Illustration of the \textit{memory effect} when skipping the first convolutional layer.}
\label{fig:fashion_memory_effect}
\end{figure}

\noindent\textbf{Exploitation.} Interestingly, we observed that exiting the loop prematurely at the first filters causes the inference results to match with the last correctly executed inference, as exemplified in Figure~\ref{fig:fashion_memory_effect}.  This \textit{memory effect} can be exploited in critical applications, such as the authentication of an unauthorized person after one with the appropriate permissions. As this type of attack leaves no trace or fault in the circuit and will be overwritten by correct data after the next prediction, it can be difficult to detect, except for monitoring the process time (e.g. by using an instruction counter). This attack is possible because the result of the convolution layer calculations is stored in RAM. Therefore, to prevent such attacks, it is compulsory to perform a RAM memory reset between two inferences to clear the stored results.

\subsection{Targeting the bias values}
\noindent\textbf{Experiments.} The first dense layers in the model contains biases that can be modified to alter the inference results. By modifying the \texttt{store} instruction that initializes the biases, we observed significant corruption of bias values resulting in mispredictions in our simulations. Specifically, we wrote an address value to the register instead of the bias so that the bias takes a significantly higher value. 
Fault injections were performed using laser injection and the value is similar to the simulation. Although the induced fault differs with EMFI, it results in bias values that are different from the initial values. This has significant effects on the inference results as shown in Table~\ref{tab_bias}. The accuracy is only detailed for the first 4 biases but they are comparable for all 24 neurons. 

\begin{figure*}[t]
\centering\hspace{1cm}
\begin{subfigure}{0.31\textwidth}
\centering
    \includegraphics[width=\textwidth]{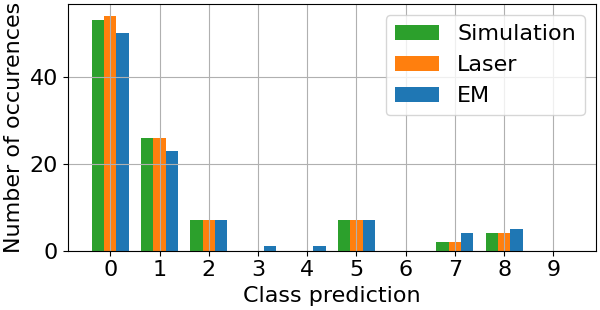}
    \caption{Fault on neuron \#0}
\end{subfigure}
\hspace{1cm}
\begin{subfigure}{0.31\textwidth}
\centering
    \includegraphics[width=\textwidth]{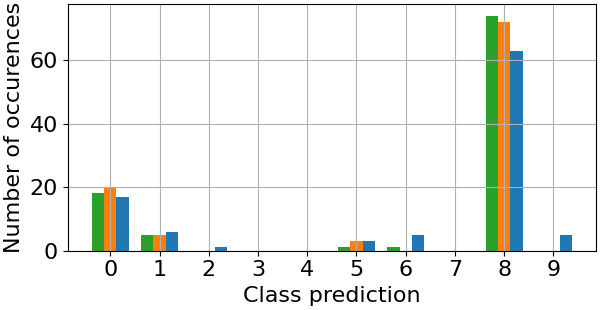}
    \caption{Fault on neuron \#1}
\end{subfigure}

\centering\hspace{1cm}
\begin{subfigure}{0.31\textwidth}
\centering
    \includegraphics[width=\textwidth]{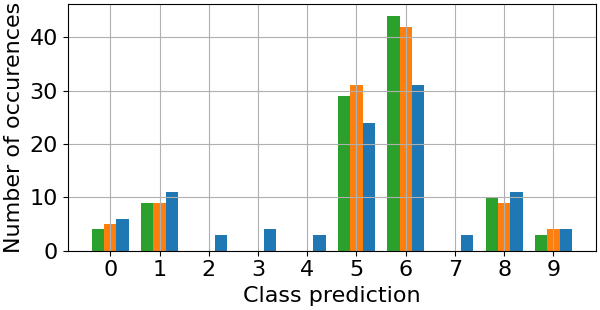}
    \caption{Fault on neuron \#2}
\end{subfigure}
\hspace{1cm}
\begin{subfigure}{0.31\textwidth}
\centering
    \includegraphics[width=\textwidth]{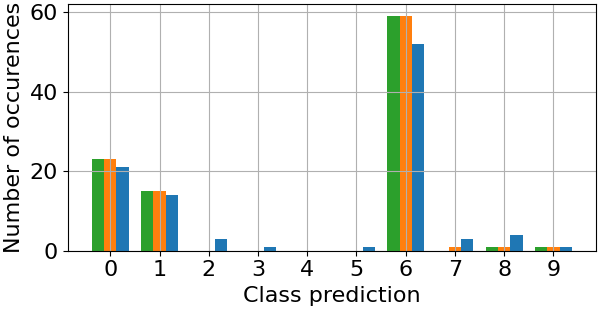}
    \caption{Fault on neuron \#3}
\end{subfigure}
\caption{Majority output inference when faulting bias on the 4 first neurons}
\label{fig:fashion_biais_histo}
\end{figure*}

\begin{table}[H]
\centering
\caption{Accuracy when faulting the bias of one of the 4 first neurons}\label{tab_bias}
\resizebox{\columnwidth}{!}{%
\begin{tabular}{x{1.6cm}x{2.5cm}x{1.8cm}x{1.8cm}}
\toprule
Neuron tested & Accuracy Simulation & Accuracy Laser & Accuracy EM \\
\midrule

0 & 38\% & 37\% & 40\% \\
1 & 26\% & 26\% & 37\% \\
2 & 40\% & 41\% & 56\% \\
3 & 28\% & 28\% & 35\% \\
\midrule
no fault & 77\% & 77\% & 77\% \\
\bottomrule
\end{tabular}
}
\end{table}

\noindent\textbf{Exploitation.} The accuracy of the model was found to be highly sensitive to a single injection, with modifications to the bias of a neuron favoring certain predictions over others, as presented in Figure~\ref{fig:fashion_biais_histo}. For example, modifying the bias of the first neuron resulted in the model mostly predicting \textit{T-shirt} (label 0) about 50 times out of 100 tests, regardless whether the simulation, electromagnetic or laser injection we used. The results for the other bias calculations are similar for the accuracy and majority inferences. These results demonstrate that it is possible to significantly bias the model predictions.

An attacker can then easily force a prediction by choosing to fault the calculation of a single bias;

To protect against this type of attack, the bias value of the dense layer can be reset if it exceeds a certain bound. In simulation on the first 4 bias calculations with bounds between -2048 and 2048 to the output value of the neuron, we obtained the same results to those without fault injection. We recovered the initial accuracy which had been reduced between 26 and 40\% with the fault injection.

\subsection{Targeting the activation function}
\label{exp_relu}
\noindent\textbf{Experiments.} We targeted the activation function of the first \textit{dense} layer composed of 24 neurons (with instruction skip faults).
For this layer, the activation function is ReLU  ($\sigma(x)=\max(0;x)$) which is generally used in most  state-of-the-art deep convolutional network models. Fig.~\ref{fig:relu}~(a) presents the assembly code of ReLU used in our NNoM implementation.

We first simulated the impact of instruction skip. It appears that it is possible to alter ReLU in two different ways: 
\begin{itemize}
    \item force a reset (target the blue instructions in Fig.~\ref{fig:relu}~(a)): output is zero while the input is positive. Therefore, the activation is constantly null: $\sigma(x)=0$.  
    \item skip the reset (target the red instructions in Fig.~\ref{fig:relu}~(a)): the activation is turned into the identity function and is not set to zero if the input is negative: $\sigma(x)=x$.  
\end{itemize}
Interestingly, our simulations shown that skipping the reset causes more mispredictions than forcing a reset. Consequently, we focused on this kind of attack for our experiments. 

\noindent\textbf{Exploitation.} Skipping the reset of the ReLU activation function has minor impact on the accuracy of the model, as depicted in the Figure~\ref{fig:relu}~(b). Therefore, when targeting the first fully-connected layer, our experiments show that it is less effective to target the ReLu activation function than the biases if we seek to reduce the accuracy. This observation is coherent with the first experiments from Breier et.al.~\cite{breier_sigsac} that demonstrated the need of performing a significant number of faults on ReLU to alter the overall accuracy of a 5-layer MLP model (faults were injected on the penultimate layer). We observe that the results of simulation experiments are similar to those obtained with LFI, with an accuracy decreasing to 74\%. The accuracy was reduced to 61\% with EMFI where we observed a certain number of outputs equal to zero (less than those without faults) that indicates few forced resets (i.e. forcing a reset while the input is positive). This behavior did not appear in simulation or with LFI.
That highlights the fact that, even though simulations can predict the majority of behaviors, experimental studies are compulsory to accurately account for the effect of a complex fault model.

%% file: src/discussions.tex
\subsection{Comparison of injection techniques and limitations}
\input{fig/relu}
LFI is known as a very effective injection means because of a high temporal and spatial accuracy~\cite{yuce2018fault}. However, contrary to EMFI, silicon has to be visible to perform LFI, therefore it is necessary to decapsulate the components, a delicate step that complicates the implementation of the attack. Another practical point between LFI and EMI is the cost of the characterization environment, significantly higher for LFI with approximately 100k\,€ compared to 30k\,€ for the EMFI bench. 
Experimentally, the search for the sensitive zone was more tedious in EMFI as we encountered many freezes and restarts of the target device, which was less common with the laser pulses. Moreover, our fault model being a single instruction skip, it was straightforward to simulate the impact of such faults on the inference process and compare the match between simulation outputs and observed ones with real injections. We observed a higher similarity between simulations and LFI than with EMFI (typically for the experiments on ReLU in~\ref{exp_relu}). This difference is explained by a lower repeatability of EMFI due to its lower spatial accuracy compared to laser. 

A classical limitation of our security characterization is related to the synchronization of the injections: dedicated instructions inserted in our test programs activated an output of the target that was used to trigger our fault injectors after a programmable delay (i.e. our experiments were performed in a white-box setting).
However, since the implementation code includes many loops that generate an electromagnetic leak with a particular and detectable signature, one can consider the use of a device to temporally trigger the injections, such as presented in \cite{smarttrigger}. We keep the use of such techniques for further experiments. 

\subsection{Neural network implementation}
Our experiments are based on a classical CNN model deployed thanks to the NNoM platform, which has the advantage of being open source, covering classical types of layer and reaching interesting inference performance. Further works would investigate how different implementations (e.g., MCUNet~\cite{lin2020mcunet}) behave when exposed to instruction skips even for challenging black-box tools (e.g., STMCubeMX.AI).

\noindent\textbf{Target a single convolutional kernel.} According to what we observed with the NNoM implementation, we can highlight some interesting outcome that pave the way to further analysis of different types of implementations for the most critical functional structures of the inference (and then lead to more robust neural network inference implementations). For example, we simulated an implementation that allows for the non-execution of a single filter in the convolution layer, rather than producing a premature exit as presented in section~\ref{exp_conv}. This attack is also possible on the NNoM implementation by replaying the instruction that increments the loop counter. As a result, an adversary could skip only one filter and then exclude one channel from the resulting features map. Our results show that the accuracy can be reduced by up to 20\% depending on the targeted filter. 
This significant drop of accuracy is consistent with what observed with only few bit-flips with the BFA~\cite{rakin2019bit} that usually result in \textit{turning off} an initially important kernel. 

\noindent\textbf{Using CMSIS-NN.} 
With NNoM, it is possible to use as backend the CMSIS-NN~\cite{lai2018cmsis} library from ARM\footnote{also used in STMCubeMX.AI from STMicroelectronics} and, in that case, the implementations may differ. Although our results were obtained without using the CMSIS-NN library, the bias loading code remains the same, making the attack transferable. For the convolution operation, CMSIS-NN uses the \texttt{im2col} algorithm~\cite{lai2018cmsis} that transforms the input image and the set of filters in a new matrix representation so that the convolution is processed through efficient matrix multiplications. We performed a first set of simulation tests on an implementation using the CMSIS-NN library that shown that a single instruction skip in the loop over the dimension of the output tensor (i.e. the number of kernels) of the first convolutional layer causes a forced output: we obtained the label 0 (T-shirt) for 98 inputs over 100. This behavior is also highly critical and may be exploited by an adversary to impact the integrity of the model. Thus, further experiments are necessary to analyze potential vulnerabilities to the \texttt{im2col} implementation.  

\subsection{Protections and Exploitation for confidentiality concern}
As a first step, we focused our experiments on the direct impact of faults on the model performance (here, the standard accuracy), i.e. we mainly focus our work on task-integrity purpose. However, recent milestone works such as~\cite{rakin2022deepsteal} demonstrated how fault injection techniques can be exploited to leak critical information about a model that may help an adversary for a model extraction attack. An interesting future work is to analyze what kind of information about the parameters can be revealed by one or several instruction skips. A first insight is that skipping filters will give important information about the importance of these filters for the prediction. This is what we observed with our experiments on the first convolutional layer (section~\ref{exp_conv})  since we highlighted the fact that exiting the loop from the $17^{th}$ kernel has a limited effect on the accuracy, meaning that most of the most important kernels are the first ones. Therefore, an adversary may put his effort only on recovering a small part of the parameters which can significantly facilitate the attack.     

Many software countermeasures have been proposed by the hardware security community to protect critical algorithms (e.g., cryptographic modules)~\cite{barenghi2010countermeasures} and we mentioned some obvious implementation advice to reduce the impact of the faults we performed in section~\ref{exp_section}. However, the main challenge in terms of defense, is the length of the inference code that contains many loops. Therefore, many protections based on local verification (including ones relying on redundancy) can lead to prohibitive additional costs and significantly reduce the performance of the system. Promising defense schemes encompass the protections based on CFI (Control Flow Integrity) that aims at checking a program execution flow and detecting potential alteration~\cite{zgheib2022cfi, de2017sofia}.

%% file: fig/relu.tex
\begin{figure*}[h!]
     \centering
     \begin{subfigure}[b]{0.45\textwidth}
         \centering
         \begin{code}[commandchars=\\\{\},fontsize=\scriptsize]
\textcolor{Green}{// if (data[i] < 0) \{}
ldr   r2, [r7, #4]
ldr   r3, [r7, #12]
\textcolor{Red}{add  r3, r2}
\textcolor{Red}{ldrsb.w r3, [r3]}
\textcolor{Red}{cmp  r3, #0}
\textcolor{RoyalBlue}{bge.n  80042ee }
\textcolor{Green}{//        data[i] = 0;}
\textcolor{RoyalBlue}{ldr r2, [r7, #4]}
\textcolor{Red}{ldr  r3, [r7, #12]}
\textcolor{Red}{add  r3, r2}
\textcolor{Red}{movs r2, #0}
\textcolor{Red}{strb r2, [r3, #0]}
\textcolor{Green}{//     \}}
\end{code}
         \caption{ReLU assembly code.}
     \end{subfigure}
     \hfill
     \begin{subfigure}[b]{0.42\textwidth}
         \centering
         \includegraphics[width=\textwidth]{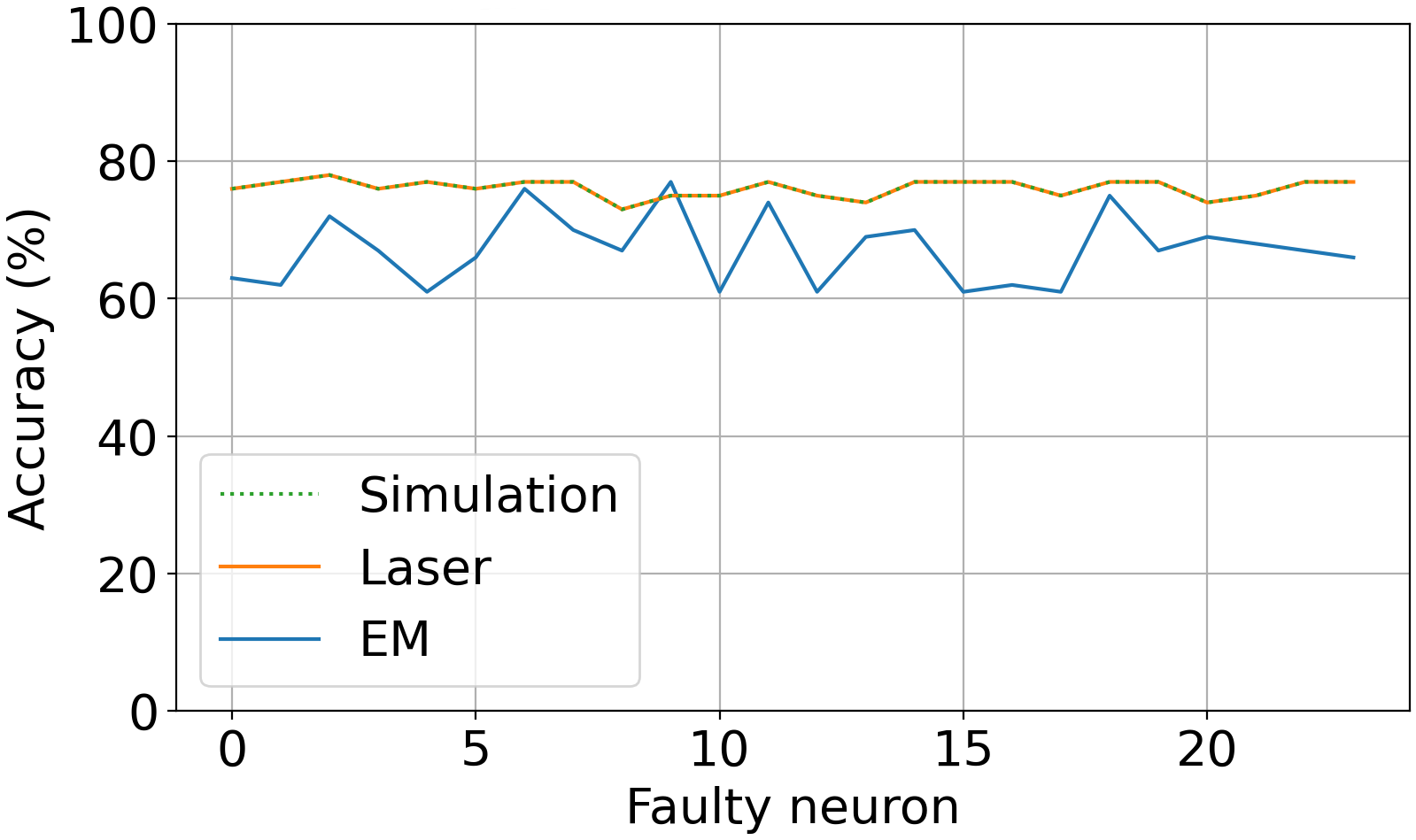}
         \caption{Impact on accuracy (Simulation, LFI, EMI)}
     \end{subfigure}
\caption{Target ReLU activation function. (Left) Assembly code. (Right) Impact on accuracy (simulation, LFI, EMI, laser and simulation curves overlap) when skipping the zeroing of the ReLU activation function.}
        \label{fig:relu}
\end{figure*}

%% file: src/conclusion.tex
We investigated the effect of single instruction skip fault attacks on a neural network model embedded in an ARM Cortex-M4 microcontroller.
We used two standard powerful injection means, laser and electromagnetic injection, as well as simulations.  We identified several vulnerabilities at different positions of the model architecture. More particularly, it is possible to prematurely exit the loop over the convolutional filters, leading to incorrect predictions, and even to a so-called memory effect if the whole convolution loop is skipped (if so, the faulted inference outputs the prediction of the previous one). Additionally, we demonstrated that instruction skips can alter the bias computation in fully-connected layer that may force the output prediction towards a chosen label. In a context of critical security concerns related to the large-scale deployment of AI systems, with upcoming regulation and certification actions, these results (the first with such an experimental scope) highlight the urgent need to properly evaluate the intrinsic robustness of embedded models and pave the way to further analysis to cover more models types, devices as well as assess the relevance of state-of-the-art protections for embedded machine learning inference programs.